\documentclass[twocolumn,amssymb,aps,epsfig]{revtex4}
\begin{document}
\draft
\author{ Georges ~Bouzerar$^{1}$ and Richard ~Bouzerar$^{2}$}
\address{
$^{1}$ Institut N\'eel MCBT CNRS 25 Avenue des Martyrs BP 156
38042 Grenoble cedex 9, France\\
$^{2}$ LPMC, Universit\'e de Picardie 33 rue Saint Leu
80000 Amiens, France \\
}
\parbox{14cm}{\rm
{\bf Comment on ``Large-Scale Monte Carlo Study of a Realistic Lattice Model for $Ga_{1-x}Mn_{x}As$''}
\medskip
}
\maketitle

Yildirim et al. have performed a Monte Carlo (MC) study of the ferromagnetism in  $Ga_{1-x}Mn_{x}As$ \cite{yildirim}. In this study, based on the dilute Kondo lattice model, a realistic band structure for the host is included. We argue that their model is inappropriate for GaMnAs. (i) The $p-d$ coupling used in their calculations is in fact too small and does not correspond to the realistic value of $1.2 ~eV$.(ii) A realistic value leads, in contrast, to very large Curie temperatures ($T_{C}$) inconsistent with the measured ones.(iii) Their MC study disagrees and contradicts previous ab-initio calculations.(iv) The experimental data plotted in Fig.1b of \cite{yildirim} are incorrectly used: $T_{C}$ were measured with samples of {\it different} Mn concentration whereas their calculations performed at fixed density.

$J$ is often denoted $N \beta$ in the literature, especially in ab-initio based studies \cite{Larson,Okabayashi,Sanvito}. It is well known that $N \beta \approx 1 ~eV$ in most of the II-VI materials as ZnMnTe, CdMnTe \cite{Larson}. Its amplitude in GaMnAs was often discussed and a value ranging from 1.2 to $3.5~eV$ 
(e.g. $ 2.5 \le JS  \le 8.8$ since $Mn^{2+}$ spin is $S=5/2$) was proposed \cite{Omiya,Matsukura,Szcytko}. However, it is now well accepted that a realistic value of $N \beta$ is close to $1.2 ~eV$ \cite{Okabayashi}, thus $JS \approx 3 ~eV$ (see the expression of $N \beta$ in \cite{Okabayashi}) should be used in realistic model studies. Although, Yildirim et al. quote this value for J, their calculations are done for $J S=1.2~eV$ since $S=1$ was used. Thus, their coupling is about 3 times smaller than the realistic value. As a consequence, according to Fig.1a of \cite{yildirim} the realistic $J$ would lead to $T_{C} \approx 700 ~K$ (5 times larger than measured). On the basis that a realistic band structure was already taken, an agreement with experimental measurement (at $x=0.085$) can be achieved within their model only if a {\it non realistic} small value of $N \beta$ is assumed. For this value, it is not clear that they could reproduce $T_{C}$ as a function of $Mn^{2+}$ as done within ab-initio calculations \cite{Lars,Georges-epl,Zunger}. Additionally, from the ab-initio studies (sophisticated realistic calculations) it is clearly seen that GaMnAs exhibits a preformed impurity band \cite{Josef,Sanvito}, $Mn^{2+}$ affects significantly the host band. For the small $J S$ used in \cite{yildirim} such effects are absent as seen in Fig.4 where the DOS at $J S =1.2 ~eV $ is almost unchanged with respect to the host DOS. Thus their $J$ corresponds to the perturbative regime (RKKY). In this regime, it is expected that the couplings may develop an RKKY tail that could eventually suppress the $T_{C}$ \cite{Richard-epl}.
Because of the very small clusters used in the MC simulations and insufficient averaging (5 configurations only), their calculated $T_{C}$ may not be reliable. They have used systems with typically only 20 spins. As seen, for example in \cite{Zunger}, even with much larger systems $T_{C}$ fluctuates already significantly from sample to sample. 

In Fig.1b of \cite{yildirim} a comparison between the calculated $T_{C}$ as a function of hole density and experimental data is shown. By reading carefully ref.\cite{exp-data} from which the experimental data were taken, one finds that the samples have different Mn concentration. Only the samples with the highest $T_{C} \approx 140 K$  have $x \approx 0.085$. The comparison with these data cannot be done, ($T_C$ also depends on $x$). Thus a comparison would require the knowledge of the total $Mn^{2+}$ and hole densities \cite{bouzerar-compens}. Hence, the agreement observed in Fig. 1b of ref.\cite{yildirim} is surprising and accidental. In addition as mentioned above, because their calculations are performed in the weak coupling regime, the stability region expected for the ferromagnetism should be smaller (the RKKY oscillations of the Mn-Mn exchange should suppress the ferromagnetic phase), in their case the $T_{C}$ increases almost linearly up to large hole density, this is in contrast with the results found from ab-initio based studies for GaMnAs \cite{Georges-epl}.
We conclude that their model is inappropriate for GaMnAs


\begin{references}
\bibitem{yildirim} Y. Yildirim, G. Alvarez, A. Moreo and E. Dagotto  Phys. Rev. Lett. {\bf 99}, 057207 (2007).
\bibitem{Larson} B. E. Larson et al., Phys. Rev. B {\bf 37}, 4137 (1988).
\bibitem{Okabayashi} J. Okabayashi et al., Phys. Rev. B {\bf 58}, R4211 (1998).
\bibitem{Sanvito} M. Wierzbowska, D. Sanchez-Portal, S. Sanvito, Phys. Rev. B {\bf 70},235209 (2004).
\bibitem{Omiya}T. Omiya, et al. Physica E (Amsterdam) {\bf 7}, 976 (2000).
\bibitem{Matsukura}F. Matsukura, H. Ohno, A. Shen, and Y. Sugawara, Phys. Rev. B {\bf 57}, R2037 (1998).
\bibitem{Szcytko}J. Szczytko et al., Solid State Commun. {\bf 99}, 927 (1996). 
\bibitem{Lars} L. Bergqvist et al., Phys. Rev. Lett. {\bf 93}, 137202 (2004).
\bibitem{Georges-epl} G. Bouzerar, T. Ziman, and J. Kudrnovsky, Europhys. Lett. {\bf 69}, 812 (2005).
\bibitem{Zunger}A. Franceschetti et al., Phys. Rev. Lett. {\bf 97}, 047202 (2006)
\bibitem{Josef}J. Kudrnovsky et al., Phys. Rev. B {\bf 69}, 115208 (2004)
\bibitem{Richard-epl} R. Bouzerar, G. Bouzerar, and T. Ziman, Europhys. Lett. {\bf 78}, 67003 (2007), condmat0607740.
\bibitem{exp-data} K.C. Ku et al. Appl. Phys. lett. {\bf 82}, 2302 (2003). 
\bibitem{bouzerar-compens}G. Bouzerar,J. Kudrnovsky and T. Ziman, Phys. Rev. B {\bf 72},125207 (2005).
\end{references}
\end{document}